\documentclass[aps,pra,twocolumn,showpacs,superscriptaddress,groupedaddress]{revtex4}  % for review and submission
\usepackage{graphicx}  % needed for figures
\usepackage{dcolumn}   % needed for some tables
\usepackage{bm}        % for math
\usepackage{amssymb}

\newcommand{\ex}[1]{\langle #1 \rangle}

\newcommand{\ket}[1]{| #1 \rangle}
\newcommand{\bra}[1]{\langle #1 |}

\newcommand{\beq}{\begin{eqnarray}}
\newcommand{\eeq}{\end{eqnarray}}

\begin{document}

%\title{Quantum State Estimation of a Double Quantum Dot}
%\title{On the Confidence of Quantum State Estimation }

%\renewcommand{\thefigure}{S\arabic{figure}}
%\setcounter{figure}{0}
%\renewcommand{\theequation}{S\arabic{equation}}
%\setcounter{equation}{0}
%\renewcommand{\thetable}{S\arabic{table}}
%\setcounter{table}{0}

\title{Confidence and Backaction in the Quantum Filter Equation}

\author{Wei Cui}\email{wcui@riken.jp}
\address{Advanced Science Institute, RIKEN, Wako-shi,
Saitama 351-0198, Japan}

\author{Neill Lambert}\email{nwlambert@riken.jp}
\address{Advanced Science Institute, RIKEN, Wako-shi,
Saitama 351-0198, Japan}

\author{Yukihiro Ota}
\address{Advanced Science Institute, RIKEN, Wako-shi,
Saitama 351-0198, Japan}

\author{Xin-You L\"{u}}
\address{Advanced Science Institute, RIKEN, Wako-shi,
Saitama 351-0198, Japan}

\author{Z. -L. Xiang}
\address{Advanced Science Institute, RIKEN, Wako-shi,
Saitama 351-0198, Japan}
\address{Department of Physics, State Key Laboratory of
Surface Physics, Fudan University, Shanghai 200433, China}

\author{J. Q. You}
\address{Department of Physics, State Key Laboratory of
Surface Physics, Fudan University, Shanghai 200433, China}
\address{Beijing Computational Science Research Center, Beijing 100084, China}
\address{Advanced Science Institute, RIKEN, Wako-shi,
Saitama 351-0198, Japan}

\author{Franco Nori}
\address{Advanced Science Institute, RIKEN, Wako-shi,
Saitama 351-0198, Japan}
\address{Physics Department, The University of Michigan,
Ann Arbor, Michigan 48109-1040, USA}

\date{\today}

\begin{abstract}
We study the confidence and backaction of state reconstruction
based on a continuous weak measurement and the quantum filter
equation. As a physical example we use the traditional model of a
double quantum dot being continuously monitored by a quantum point
contact. We examine the confidence of the estimate of a state
constructed from the measurement record, and the effect of
backaction of that measurement on that state.  Finally, in the
case of general measurements we show that using the relative
entropy as a measure of confidence allows us to define the lower
bound on the confidence as a type of quantum discord.

\end{abstract}
 \maketitle

\section{Introduction}

Recently there has been a great deal of activity on the topic of
``weak'' quantum
measurements~\cite{Wiseman2010,Alicki05,Kofman11,Ashhab,Ashhab09,Croke06,Braginsky}
in both mesoscopic~\cite{Korotkov99,Korotkov01,Korotkov06} and
macroscopic systems~\cite{Buluta11,Macroscopic,You05,Johansson}.
In contrast to projective measurement, weak measurements only
perturb the system
 slightly, but in turn can only  provide limited information.
According to
 the theory of open quantum systems,  both the evolution of the
quantum
 state and its decoherence depend on the system-apparatus
 coupling strength and the basis in which the measurement system is measured.
One of the advantages of such a measurement is that, given a weak
continuous measurement record one can reconstruct the quantum
system state during its evolution. One particularly interesting
approach, which we apply and investigate here, is the ``quantum
filter equation'' as pioneered in the quantum limit by Belavkin
\cite{Belavkin1999} and others \cite{Carmichael}.

The quantum filter equation has been shown to be a powerful method
for state reconstruction, and is fairly robust in terms of the
resolution needed in describing the measurement record.  For
example, in recent work  \cite{Ralph} it has been shown that the
continuous ``analogue'' measurement record can be reduced to a
``one-bit record'' and still the filter equation can efficiently
produce an estimate (or purification) of the system state. Similarly
it has been shown that feedback control can be used to enhance the
speed of the state estimation \cite{ Filter, Chase, Zhang}, and that
it can be further optimized when combined with a kind of process
tomography \cite{Deutsch}.  However, successful application of the
filter equation requires accurate knowledge of the evolution, both
coherent and incoherent, that the monitored system undergoes
\cite{Hill2011}.

Here we first investigate how this state estimation method can be
used by considering both how confidently \cite{Croke2006} an estimated state
deduced from the measurement record reflects the actual state of
the system, and how the effect of backaction changes the state.
Here the ``actual state of the system'' means the system under the
influence of the back-action  \cite{ backaction,Buscemi, Luo} of
the measurement apparatus, not the initial prepared state.  In
other words, we quantify the confidence of the state-estimation
process independently from the overall fidelity of the
measurement.  Thus, we focus on understanding the state reconstructed using the filter equation, without actively undoing the backaction or employing feedback. We separately define the overall accuracy of the measurement with another quantity or distance, which we term the ``epitome''. Second, we introduce a new ensemble averaged version of the filter equation which enables a more efficient numerical method (in theoretical treatments) with which to consider state estimation via the filter equation. Third, in the final section we consider the more general situation of an asymptotic positive operator valued measure (POVM), and show that in a limiting case the confidence has a lower bound set by the quantum discord.

Quantities like the confidence have
been commonly employed in investigating information gain with projective
and general measurements \cite{Banazek, Fuchs}.  The behavior of
these quantities in the context of weak continuous measurement has
not been studied in great detail as of yet, though the concept of
information gain and measurement disturbance is well understood
\cite{Jacobs,Jacobs2}. In addition, we point out that our approach
here is different from that used in some other works. For example,
here we are concerned primarily with the optimal measurement of an
unknown state by refining our state of knowledge, whereas in some
other approaches the goal is primarily manipulating (or purifying)
one's state of knowledge of a given quantum system, and the
initial unknown quantum state is unimportant~\cite{Jacobs2}.

The model we use here is based on the continuous measurement of a
double quantum dot (DQD) charge state using a quantum point
contact (QPC) \cite{Korotkov99,DQD, Petta, Lambert2010}. In this
situation the tunnelling barrier of the QPC is modulated by the
charge in the nearby DQD, and produces a continuous measurement
record which can be used in the filter equation.  However, the
approach is quite general, and recent applications have also
arisen in circuit-QED
systems~\cite{Johansson,Korotcqed}. %In this article we first
%outline how to define both confidence and backaction in this
%context.  We then give the details of our model of the DQD and QPC
%system, and illustrate the type of results the filter equation can
%give.  Finally, we outline the meaning of the backaction and
%confidence in the context of a general positive operator valued
%measure (POVM).

This article is organized as follows: In Sec.~II we provide a
general definition of the confidence, backaction and epitome. In Sec.~III we
introduce the model for weak measurement of a DQD, and show the
results given by the quantum filter equation. In Sec.~IV we
speculate about a possible filter equation based on
ensemble-averaged measurements.
Finally, in Sec.~V we show that
using the relative entropy as a measure of confidence allows us to
define the lower bound on the confidence as the quantum discord.

\begin{figure}
\includegraphics[width=3.1in]{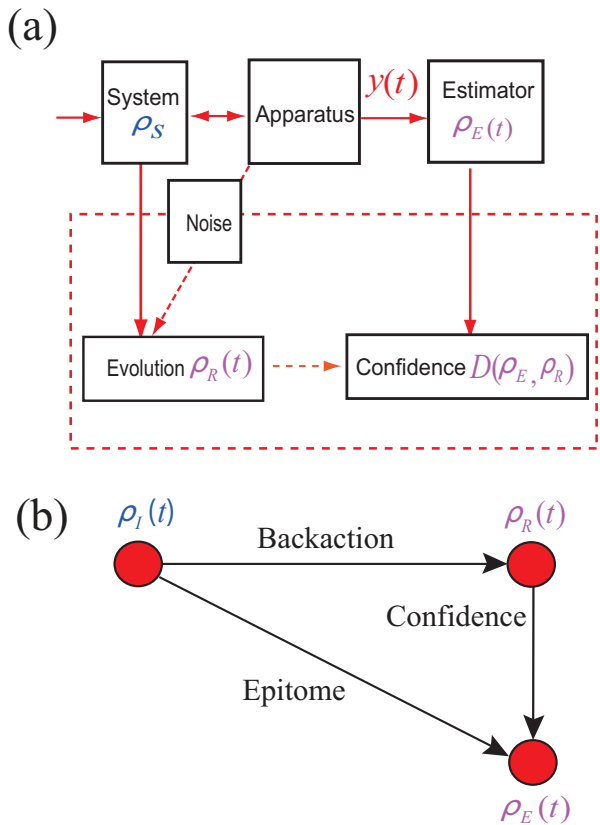}
\caption{(Color online) (a) Schematic of the components of the
quantum state estimation using the filter equation. Here, $\rho_s$
is the system initial state, $\rho_R(t)$ is the evolution of the
initial state of the system in the measurement-induced noises,
$\rho_E(t)$ is the estimation of the system state, and
$\mathcal{D}[\rho_E,\rho_R]$ is the estimation confidence. (b) shows
a diagrammatic representation of the confidence $\mathcal{C}$ and
backaction $\mathcal{B}$ as the distance between the various states.
Note that when we use the relative entropy as a distance measure
these quantities are asymmetric (as represented by the one-way
arrows).}
\end{figure}

\section{Definition of Confidence and Backaction}
Consider a quantum
system $\mathcal{S}$ which interacts with a measurement apparatus
$\mathcal{A}$. When implementing a general quantum measurement
 $\mathcal{M}=\{\Pi_i\}$, our knowledge of the system state is based only on the measured apparatus
output $y(t)$. We denote our estimate of the system state as
$\rho_E(t)$. The initial state of the system is defined as
$\rho_s(0)$, and $\rho_R(t)$ is defined as the
 evolution of the initial state
 of the system in the measurement induced noise, i.e., what we can think of as
 the actual
state of the system.  We define the ideal state of the system, i.e.,
its evolution if it were not affected by the measurement apparatus
at all, as $\rho_I(t)$, i.e., the entirely coherent evolution of
$\rho_s(0)$.

\textbf{ Definition}: %Let $S$ be a state space over quantum system
%$\mathcal{S}$ with norm $\|\cdot\|:S\rightarrow \mathbb{R}$.
The quantum state estimation confidence is defined as
  \begin{eqnarray}
\mathcal{C}\equiv D[\rho_E,\rho_R],
\end{eqnarray}
where $D[\cdot]$ is some appropriate distance measure. Generally
speaking, the smaller $\mathcal{C}$ is the more confident we are of
the estimated state, and $\mathcal{C}=0$ if and only if
$\rho_E=\rho_R$, which means that the estimated state is totally
confident. Similarly we define the backaction by
  \begin{eqnarray}
\mathcal{B}\equiv D[\rho_I,\rho_R],
\end{eqnarray}
so that $\mathcal{B}=0$ implies no backaction.  Finally, we define the overall accuracy of the measurement
with the ``epitome'',
  \begin{eqnarray}
\mathcal{E}\equiv D[\rho_I,\rho_E],
\end{eqnarray}
which naturally completes the triangle in Fig. 1.  In the treatment
that follows of course we have full knowledge of all these states at
all times, and can thus in a theoretical sense identify the optimal
parameters that minimize these
quantities. % In an experiment one could in principle prepare
%$\rho_s(0)$
%The measurement induced noises can in principle be measured and
%fed into the numerical simulation, and thus while in general
%$\rho_s(\Pi_i)$ represents only a ``best guess'' for the
%ideal state, it is a practical one.  Even in the ideal case, given
%perfect knowledge of the system state throughout its evolution,
%our estimate of the state may not coincide with it, so we define
% the quantum state estimation confidence to quantify the difference.

There is some freedom in choosing an appropriate measure for
$\mathcal{C}$, $\mathcal{B}$, and $\mathcal{E}$.  Here we explicitly consider both
the fidelity and the relative entropy \cite{Modi}. The fidelity is
commonly used to measure the effectiveness of the filter equation,
and is defined as
 \beq
F=1-\mathcal{C}=|(\sigma^{1/2}\rho\sigma^{1/2})^{1/2}|^2.\eeq We
define the confidence as $\mathcal{C} = 1-F$ in this case, to match
our definition of a distance measure, so that high fidelity implies
$\mathcal{C}=0$ (and the same with the other measures).  However since the fidelity is a pseudo-distance
this lacks some characteristics of a true measure.
 In the case of the relative entropy we
define, \beq \mathcal{C}=S(\rho_R|| \rho_E),\quad
\mathcal{B}=S(\rho_I||\rho_R),\quad
\mathcal{E}=S(\rho_I||\rho_E)\eeq where \beq S(\sigma||\rho) = -
\mathrm{Tr} \sigma \ln \rho - S(\sigma).\eeq The relative entropy
can be seen as a distance measure, though as it is asymmetric in
$\sigma \leftrightarrow \rho$ it is technically not. In fact the
ordering here is important;  with the inverse ordering the
backaction $B\rightarrow \infty$ as $\rho_I$ is a pure state in
this example. Using the relative entropy allows us to make a more
direct connection to a general POVM description of a weak continuous measurement in the final
section of this work.

\begin{figure}
\includegraphics[width=2.5in]{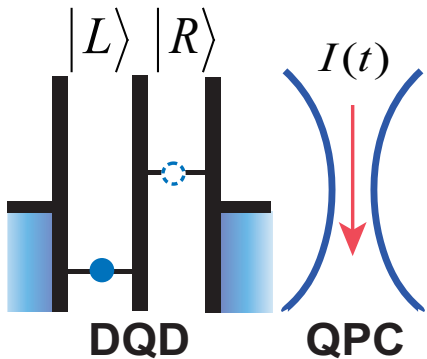}
\caption{(Color online) Schematic of a QPC used for measuring the electron states yields backaction
on the DQD.}
\end{figure}

\section{Continuous Weak measurement of a Double Quantum Dot} We now present the
specific details of the DQD and QPC system.  A DQD consists of a dot
$L$, connected to an emitter, and dot $R$, connected to a collector
\cite{DQD,Ouyang10,Petersson10}.  As is typical, we assume that the
DQD is in the strong Coulomb regime such that only one electron is
allowed in the whole DQD.  Here we assume the DQD is prepared in a
single electron state, then isolated from the emitter and collector
reservoirs via manipulation of gate voltages.  The two single-dot
states are denoted by $ |L\rangle$, and $|R\rangle$. The Hamiltonian
of the DQD can be written by \beq H_{{\rm DQD}}=\Omega\sigma_x/2 +
\epsilon \sigma_z,\eeq with $\sigma_x= \ket{L}\bra{R} +
\ket{R}\bra{L}$, $\sigma_z =\ket{L}\bra{L} - \ket{R}\bra{R}$, $\epsilon$ is the level
splitting between the two single-dot states and $\Omega$ is the
coherent tunnelling amplitude betweens the two dots. The
nearby QPC has the Hamiltonian $H_{{\rm
QPC}}=\sum_k\epsilon_{1k}a_{1k}^{\dag}a_{1k}+
\sum_{q}\epsilon_{2q}a_{2q}^{\dag}a_{2q}$, and the interaction
Hamiltonian $H_{I}=\sum_{k,q}(\chi-\chi_L|L\rangle\langle
L|-\chi_R|R\rangle\langle R|)
(a_{1k}^{\dag}a_{2q}+a_{2q}^{\dag}a_{1k})$, which is modulated by
the electron states of the DQD. Here $\chi$ is the tunneling amplitude of an
isolated QPC, $\chi_L (\chi_R)$ gives the variation in the tunneling amplitude when the
extra electron stays on the left (right) dot, and $a_{1k}  (a_{2q})$ denotes the
electron annihilation operator for the source (drain) of the QPC.
Because the height of the tunneling barrier in the QPC depends on
the electron states of the DQD, it is expected that the measured
current of the QPC will vary with the DQD states.

In this simple model, the current shot noise \cite{Aguado04} of the
QPC acts as a noise source. In the low-temperature limit with
$k_BT\ll\hbar\omega$, the noise spectral density takes the form
\cite{Aguado00,Aguado04,Gustavsson07,Petersson10} \beq
 J_I(\omega)=\frac{4}{R_K}D(1-D)\frac{(eV_{\rm QPC}-\hbar\omega)}{[1-
 \exp\{-(eV_{\rm QPC}-\hbar\omega)
/k_BT\}]},\eeq where $R_k=h/e^2$ is the von Klitzing constant,  $D$ is the transparency of a single channel in
the QPC, and $D$ is a function of $\chi$, $\chi_L$ and $\chi_R$ \cite{Korotkov01, Aguado00,Gustavsson07,Lesovik}. To treat the
measurement signal, or current through the QPC, as a classical
variable one must assume that the QPC evolution is much faster than
the DQD, so that only the zero-frequency component is important;
this is effectively a Markovian limit in terms of treating the QPC
backaction on the DQD \cite{Korotkov99,backaction}.

We also treat the QPC as a perfect detector. In this limit we can
define the measurement strength of the QPC as \beq
\kappa=\frac{(\Delta I)^2}{16 J(0)}.\eeq  Here, $\Delta I=I_L -
I_R$, where $I_{L} = D_{L} e^2 V/ \pi \hbar$,$I_{R} = D_{R} e^2 V/
\pi \hbar$, $D_{L} = D+\Delta D$, $D_R = D-\Delta D$, and $\Delta D$
is the change in the transmission of the QPC due to the charge state
of the DQD.

\begin{figure*}
\setlength{\abovecaptionskip}{7pt}
\includegraphics[width=2\columnwidth]{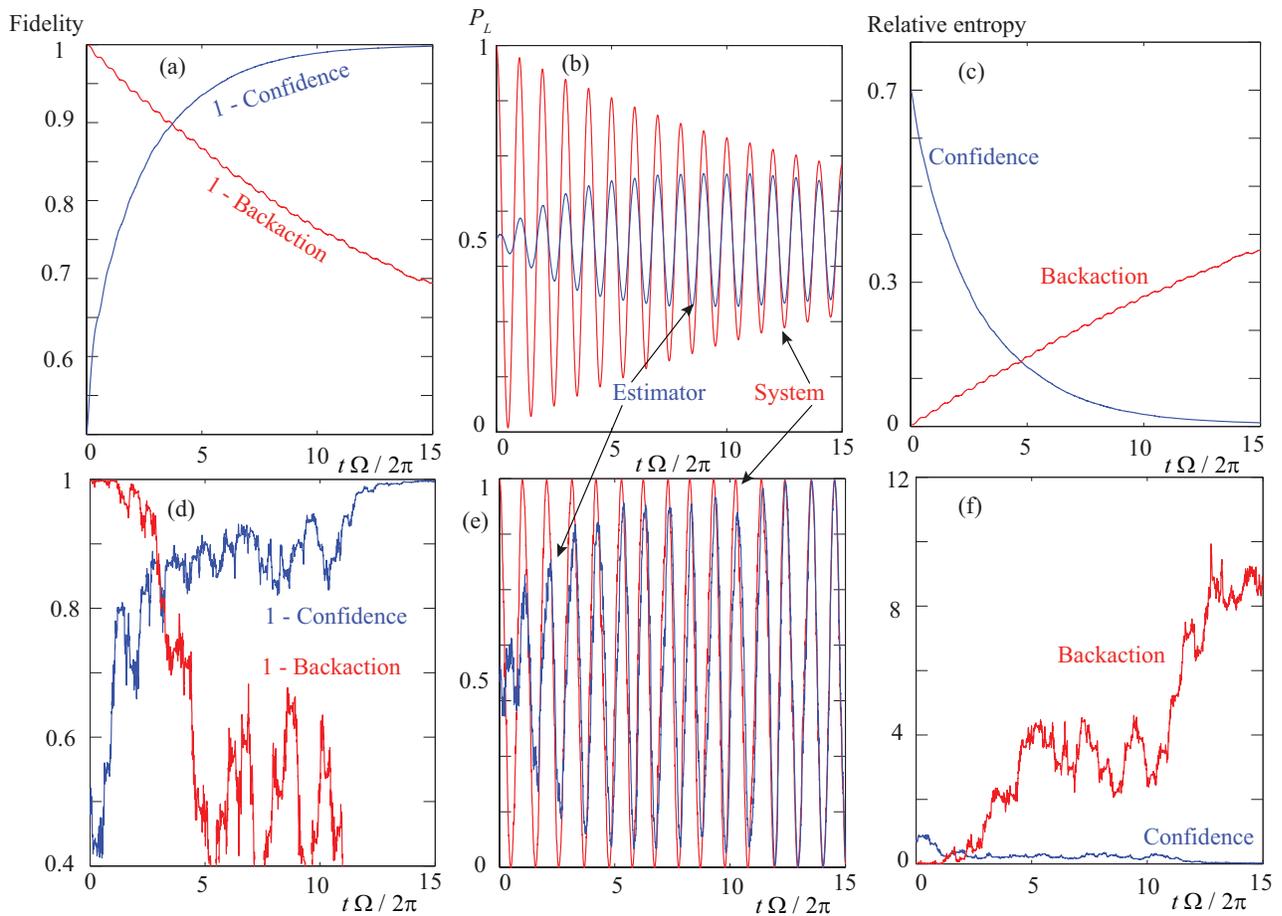}
 \caption{(Color online).  The top row of figures shows the ensemble-averaged
 behavior (over $2,000$ realizations) of
 (a) one minus the confidence (blue) and one minus the backaction (red) defined in terms of the fidelity, (b) the occupation of the left ``dot'' or state for the system,
 Eq.~(\ref{Master}) (in red) and the estimator, Eq.~(\ref{ME}), (in blue),
 and (c) the confidence (blue) and backaction (red)
 defined in terms of the relative entropy.  In all figures $\kappa = 0.005 \Omega$, and time is evolved for the equivalent of $15$ Rabi oscillations of the bare quantum dot state.  Figure (b) shows more directly how the estimator quickly oscillates in phase
 with the system state, but takes time to evolve to the same population magnitude.  The small oscillations seen in both backaction curves in (a) and (c) are typical for the definition of the backaction, and simply represent the ``nodes'' in the oscillation curve of, e.g., $P_L$, where the $\rho_I$ and $\rho_R$ states coincide.
 Figures (d)-(e) show the same quantities as (a)--(c) except just a single realization.
 The estimator state still quickly approaches the system state,
 as is typical with the filter equation.  The backaction changes in
 magnitude drastically in both (d) and (f).  There is no single-realization
 dephasing in (e) because we assume the QPC to be a perfect detector.
 This is in contrast to recent work \cite{Johansson} on circuit QED where some information
 can remain hidden due to the finite lifetime of the measurement cavity. Note, in (a) and (d) we have plotted $1-C$ and $1-B$, so that the fidelity result can be easily compared to other works investigating the effectiveness of the filter equation using fidelity.}\end{figure*}

The evolution of the real state of the DQD $\rho_R$ can be derived
using the Bayesian techniques of Korotkov
\cite{Korotkov99,Korotkov01,Korotkov06} to give the selective
stochastic master equation (SME) in Ito form,
 \begin{eqnarray} \label{Master}
\rm{d}{\rho_R}&=&-\frac{i}{\hbar}\left[H_{\rm
DQD},\rho_R\right]dt+
\kappa\mathcal{D}[\sigma_z]\,\rho_R\,dt \nonumber\\&+&\sqrt{2 \kappa}\;\mathcal{H}[\sigma_z]\;\rho_R \;dW_R,\nonumber\\
 \end{eqnarray}
 where $\kappa$ is defined above,
\beq \mathcal{H}[\sigma_z]\rho_R\equiv \sigma_z\rho_R+\rho_R
\sigma_z^{\dagger}-\mathrm{Tr}(\sigma_z\rho_R+\rho_R
\sigma_z^{\dagger})\rho_R,\eeq and the real Wiener process
 $dW_R$ satisfies $\mathrm{E}(dW_R)=0,$ $(d
W_R)^2=dt$.
 In Eq. (\ref{Master}), the super-operator $\mathcal{D}$ is
defined as \beq \mathcal{D}[a]\rho=a\rho
a^{\dag}-\frac{1}{2}(a^{\dag}a\rho+\rho a^{\dag}a).\eeq

%  The spectral density of the apparatus-induced noise can be
%experimentally evaluated, e.g., by using a multiple-pulse sequence
%\cite{Yuge11} or  dynamical decoupling \cite{Alvarez11}.
 Given that, in a general sense, the measurement
operator is $ y= \sqrt{2\kappa\hbar}\sigma_z$  the measurement
record increment at a time $t$ is, \beq \frac{dy(t)}{\sqrt{\hbar}}
&=& \sqrt{8\kappa} \ex{\sigma_z^R(t)} dt + dW_R=\frac{\Delta I
\ex{\sigma_z^R(t)}}{\sqrt{2 J(0)}}dt +dW_R.\nonumber\\ %%&\equiv&
%%\frac{d(I(t)-I_0)}{\sqrt{2J(0)}}.\nonumber
\eeq Here \beq \ex{\sigma_z^R(t)} = \mathrm{Tr}[\sigma_z
\rho_R(t)]\eeq is the instantaneous expectation value of
$\sigma_z$ at time $t$ based on the selectively evolved density
matrix $\rho_R(t)$.

This equation of motion also relies on the assumption that \beq
|I_L-I_R|=|\Delta I|\ll I_0=(I_L+I_R)/2,\eeq so that many electrons,
$N\geq(I_0/\Delta I)^2\gg1$, should pass through the QPC before one
can distinguish the quantum state. This is the weakly responding or
linear regime, and the model as we have described it is entirely
equivalent to the formulation used by Korotkov and
others~\cite{Korotkov99,Korotkov01,Korotkov06}. Also, note that for
consistency with other works on the filter equation as a state
estimation technique~\cite{Ralph,Hill2011} we describe the noise as
a Wiener process, so that the width of the Gaussian distribution
\cite{Korotkov01,Korotkov06} used to describe the weak measurement
with a QPC is absorbed into $\kappa$.

\subsection{Quantum Filter Equation}
To estimate the quantum state $\rho_E$ from the measurement output
we employ the quantum filter equation method
\cite{Shannon,Mitter,Ralph,Filter, Chase, Zhang}.
 The
evolution of the estimated state $\rho_E$ is described by the
following stochastic master equation, identical to the ``system"
one, except the measurement signal from the system evolution,
described above, determines the noise term:
 %\begin{eqnarray} \label{ME}
%&&d\rho_E=-\frac{i}{\hbar}\left[H_{\rm DQD},\rho_E\right]dt+
%\kappa\mathcal{D}[\sigma_z]\rho_E dt \nonumber\\&&+\sqrt{2 \kappa}\mathcal{H}[\sigma_z]\rho_E %\left[\frac{\Delta I}{\sqrt{2J(0)}} \left\{ \ex{\sigma_z^R(t)} - \ex{\sigma_z^E(t)}\right\}dt+dW_R\right],%\nonumber\\
 %\end{eqnarray}

 \begin{eqnarray} \label{ME}
&&d\rho_E=-\frac{i}{\hbar}\left[H_{\rm DQD},\rho_E\right]dt+
\kappa\mathcal{D}[\sigma_z]\rho_E dt \nonumber\\&&+\sqrt{2 \kappa}\mathcal{H}[\sigma_z]\rho_E \left[
\frac{dy(t)}{\sqrt{\hbar}}-\frac{\Delta I}{\sqrt{2J(0)}}\ex{\sigma_z^E(t)}
\right],
 \end{eqnarray}

The last term  is analogous to the classical \emph{innovation} in
control theory \cite{Shannon,Mitter}, i.e., the difference between
the actually measured current and the predicted current with the
estimated state.  The state estimation process involves setting
$\rho_E(0)=1/2$, and evolving under the noise determined by the
measurement record from the ``experiment'', or in this theoretical
work Eq. (\ref{Master}).

As demonstrated in Ref.~[\onlinecite{Hill2011}], even a small
error in the Hamiltonian of the above equation can induce errors
in the estimate of the state provided by the quantum filter
equation. We expect the same to be true of the estimates of the
noise spectrum of the QPC. Here our goal is to study the
efficiency and robustness of quantum state estimation via the
filter equation, so as in Ralph \textit{et al.}
[\onlinecite{Ralph}] we choose the same Hamiltonian and QPC
properties in both Eq. (\ref{ME}) and Eq. (\ref{Master}). We leave
the problem of accurately estimating the Hamiltonian and the noise
properties of the measurement in a dynamic way \cite{Hill2011} for
future work.

\begin{figure}
\setlength{\abovecaptionskip}{7pt}
\includegraphics[width=0.85\columnwidth]{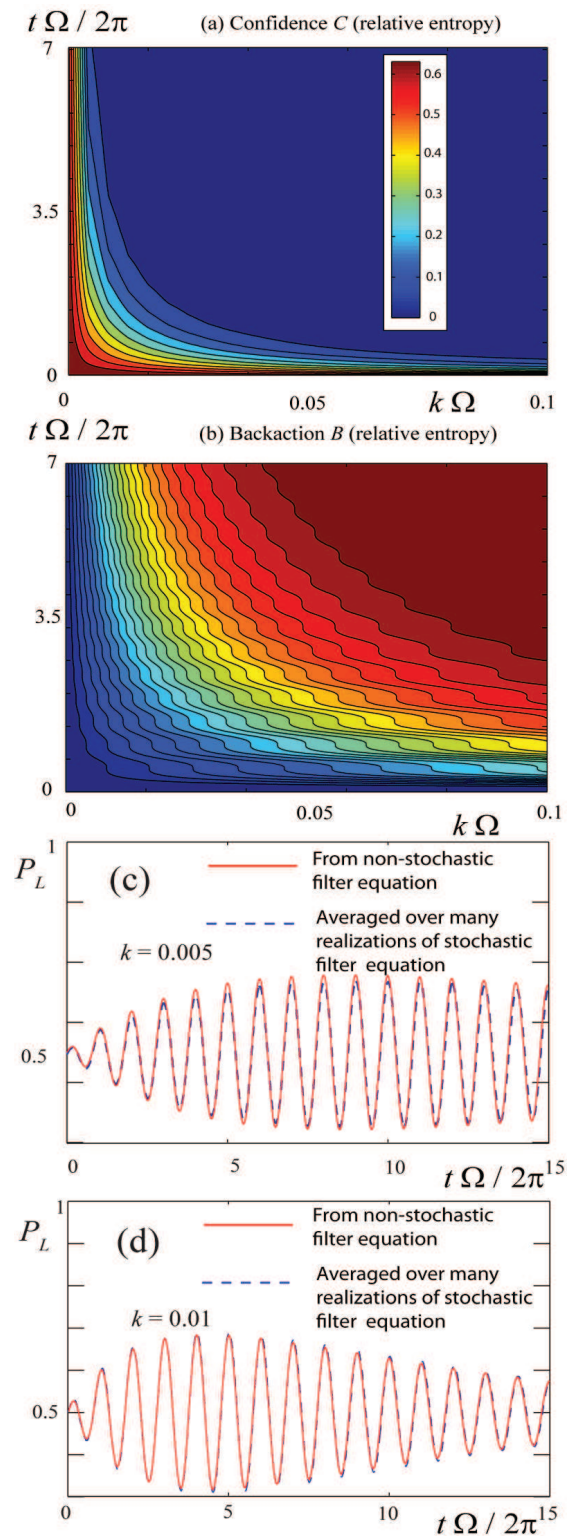}
 \caption{(Color online) Ensemble-averaged (a) confidence $C$ and
 (b) backaction $B$ in terms of the relative entropy between system and
 estimator states, as a function of the interaction time $t$ and measurement
 strength $\kappa$.  Both figures
 are derived using the ensemble-averaged equation of motion (\ref{ME2}).
 (c) and (d) show a comparison between the probability of occupation of
 the left dot $P_L(t)$ derived from averaging Eq. [\ref{ME}] over many
 realizations (dashed blue lines) to that
 given by solving Eq. [\ref{ME2}].}\label{fig4}
\end{figure}

Finally, we combine the time-evolution of these two equations
(\ref{Master}) and (\ref{ME}) to calculate the confidence of the
estimated state and backaction of the measurement using both the
fidelity and the relative entropy, as described earlier. In Fig.3,
we show numerical results for these quantities.  We will explicitly discuss the epitome in the next
section.  We set
$\rho_R(0)=\ket{L}\bra{L}$, $\rho_E(0)=1/2$, and evolve using the
standard techniques for $150000$ time steps per Rabi oscillation.
The top row of figures shows the results averaged over $2000$
realizations, while the bottom row shows just a single realization.
In these results we set $\hbar=1$, $\epsilon=0$ , $\Omega=1$ and
$\kappa=0.005 \Omega$.  Comparing to real parameters from
 \cite{Gustavsson07,Petersson10}, one could choose a strong inter-dot coupling
 of the order of
$\Omega=32$ $\mu e\mathrm{V}$, giving a timescale of $130$ ps for
the Rabi oscillations we show in Fig. 3.  This should be chosen
carefully however, to match the desired properties of the QPC
timescales (or whatever the measurement device happens to be).

 When we acquire  information
from the measurement, it of course induces significant backaction
on the system itself. Figure 3.(a) shows the confidence and
backaction in terms of the fidelity, while Fig. 3.(c) shows the
same in terms of the relative entropy.
 Both give reasonable descriptions of the distance between the estimated state and system state, and for weak measurement strength
 the confidence saturates before the backaction does.  Obviously then the trade off is
 to measure on a time scale where both the confidence is relatively high and the backaction is low.

To gain more insight on what is actually happening during the
evolution of the filter equation, Fig. 3(b) and (e)  show the
population of the left state of the dot for both the system and
estimator. In the ensemble averaged case Fig. 3(b) backaction from
the measurement dephases the state, but the estimator matches the
system state before coherent information is totally lost.  In a
single realization, 3(e), the system state does not dephase
because the QPC acts as a perfect detector.  Compare this to the
case of circuit-QED where the lifetime of the cavity can induce
dephasing on certain timescales \cite{Johansson, Korotcqed}.  The
off-diagonal elements behave in a similar fashion.

\section{A filter equation for ensemble expectation values}

Solving for the ensemble-averaged results by collating many single
realizations is sometimes an arduous numerical task, %(particularly
%for intermediate or large values of $\kappa$)
though can be useful
in the stochastic Schr\"{o}dinger form if one is modelling a
system with a large Hilbert space.  %Similarly in many experimental situations
%continuous weak measurements cannot be performed at all, or the
%individual trajectories cannot be isolated; only ensemble averaged
%results, like $\mathrm{E}[\ex{\sigma_z^R(t)}]$, are available.
In practise the system state density matrix ensemble averaged over
all measurement trajectories can be trivially calculated by
averaging Eq. (\ref{Master}), and noting $E[dW_R]=0$. This gives
the expected Lindblad equation of motion which induces the
behavior we observe in the backaction. How about the estimator
state? Averaging Eq. (\ref{ME}) is non-trivial as the individual
trajectories determined by $\ex{\sigma_z^R(t)}$ are not
statistically independent of $\rho_E(t)$.

Rather than attempt to do so we simply write down an analogy to the
quantum filter equation which depends on ensemble-averaged
quantities rather than individual trajectories. We now define
$\rho_E=\mathrm{E}[\rho_E]$,
$\ex{\sigma_z^R(t)}=\mathrm{E}[\ex{\sigma_z^R(t)}]$, and
$\ex{\sigma_z^E(t)}=\mathrm{E}[\ex{\sigma_z^E(t)}]$. Thus the term
$\mathrm{E}[\ex{\sigma_z^R(t)}]$ represents the expectation value
extracted from an ensemble averaged version of Eq.~(\ref{Master}),
i.e., the evolution of the real system under the effect of
dephasing.  By comparison to the stochastic filter equation we
consider the following nonstochastic filter equation,
\begin{eqnarray} \label{ME2}
\frac{\rm{d}\rho_E}{\rm{d}t}&=&-\frac{i}{\hbar}\left[H_{\rm
DQD},\rho_E\right]+
\kappa\mathcal{D}[\sigma_z]\rho_E \nonumber\\&+&\sqrt{2 \kappa}\mathcal{H}[\sigma_z]\rho_E \left[\frac{\Delta I}{\sqrt{2J(0)}} \left\{ \ex{\sigma_z^R(t)}  - \ex{\sigma_z^E(t)}\right\}\right].\nonumber\\
\end{eqnarray}
Solving this equation directly is computationally trivial.  To
illustrate this we plot the confidence as a function of time and
measurement strength $\kappa$ in Fig. 4. We can easily see that as
$\kappa$ increases the confidence saturates quickly, but with an
associated strong backaction, as expected.  Remarkably, if we
inspect the density matrix elements of the estimated state generated
by Eq. (\ref{ME2}) to those generated by the ensemble average over
trajectories of Eq. (\ref{ME}) they coincide closely. This is
illustrated in Fig. 4(c) and (d).  Curiously we are unable to
rigorously justify this correspondence, though one can note that Eq.
(\ref{ME2}) represents a valid solution to Eq. (\ref{ME}) for the
trajectory determined by $dW_R=0$. We also point out that the
fictitious nonlinear force in the second line of Eq. (\ref{ME2}) is
not physical, and the equation may not ensure positivity of the
density matrix $\rho_E$ at an arbitrary time. Why this works so well
in reproducing results from the ensemble averaged filter equation,
at least for this case of a single qubit measured in the $\sigma_z$
basis, is not clear, and represents a possible avenue of future
work.

\begin{figure}
\setlength{\abovecaptionskip}{7pt}
\includegraphics[width=1\columnwidth]{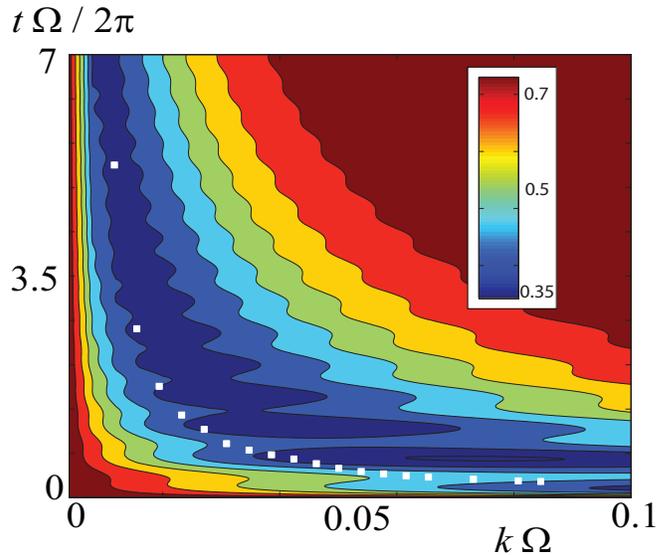}
 \caption{(Color online)  The epitome, $\mathcal{E}=S(\rho_I||\rho_E)$, as a function of the interaction time $t$ and measurement
 strength $\kappa$. This figure is
 derived using the ensemble-averaged equation of motion (\ref{ME2}).  The superimposed
 white squares indicates the line of crossing points between the confidence and backaction of figure \ref{fig4}, and naturally
 is close to the minimum in the epitome.
}
\end{figure}

Finally, in Fig. 5., we use this nonstochastic filter equation to plot the epitome, $\mathcal{E}=S(\rho_I||\rho_E)$.
We see that at some intermediate time, depending on the measurement strength, the epitome
has an optimal minima which coincides closely to the crossing point of the confidence and backaction.
Thus, as one would expect, with continous weak measurements there is an optimal time at which our estimated
state is closest to the original, unperturbed, ideal state.  In practise this optimization can be also discussed in terms
of goal programming (shown in Appendix A) \cite{Goal}. Further methods to improve the estimation,  or minimize backaction, could include feedback or other techniques from
quantum control \cite{ueda}.

\section{POVM and the discord as a bound on confidence}

In quantum theory one can describe any measurement scenario as a
positive-operator valued measure (POVM).   For example, weak
measurement is sometimes discussed in terms of a POVM on a
combined system/measurement-apparatus, where the measurement
apparatus itself is also considered to be a quantum system. To
gain a more fundamental perspective on the confidence and
backaction, as we have defined them here, we consider an
alternative measurement scheme of an asymptotic POVM.

First, we retain our definition of the initial pure system-apparatus
state $\rho_I(0)=\rho_s(0)\otimes\rho_A(0)$.  We then assume that a
measurement apparatus $A$ is allowed to interact with the system to
produce the typically entangled and correlated system-apparatus
state $\rho_{s,A}$ (we suppress any time argument here for complete
generality). We define the ``real'' state of the system then as this
combined state $ \rho_{s,A}$. Finally, we perform measurements on
$A$ associated with a POVM $\{\Pi_j^{A,\dagger}\Pi_j^A\}$, where
$\sum_j \Pi_j^{A, \dagger} \Pi_j^{A}=1$. Our estimate of the
combined system-apparatus state given by the measurement is $\rho_m
= \sum_j \Pi_j^A \rho_{R} \Pi_j^{A, \dagger}$.  Again we define the
confidence in terms of the relative entropy, so $C=S(\rho_{s,A}||
\rho_m)$.%, and $B=S(\rho_p||\rho_R)$

%Since $S(\rho_p)=0$, the backaction just describes the distance
%between the pure state $\rho_p$ and the mixed state $\rho_R$.
The relative-entropy-based confidence has an interesting lower bound
in the case when the POVM becomes a projective valued measure.
Writing,
\begin{eqnarray} C = -\mathrm{Tr} \rho_{s,A} \ln \sum_j \Pi_j^A
\rho_{s,A} \Pi_j^{A, \dagger} - S(\rho_{s,A})  \end{eqnarray} we
can substitute $\Pi_j^A =\ket{j}\bra{j}$, where $\ket{j}$ is some
distinguishable orthonormal basis describing the measurement
apparatus.  Then the confidence becomes,
%\begin{eqnarray}
%C &=&
%-\mathrm{Tr}\sum_j\ket{j}\bra{j}\rho_{s,A}\log
%\sum_j\ket{j}\bra{j}\rho_{s,A}\ket{j}\bra{j}- S(\rho_{s,A})\nonumber\\
%&=&-\mathrm{Tr}\sum_j\ket{j}\bra{j} \rho_{s,A}\ket{j}\bra{j} \log
%\sum_j\ket{j}\bra{j}\rho_{s,A}\ket{j}\bra{j}
%- S(\rho_{s,A})\nonumber\\
%&=& S(\rho_m) - S(\rho_{s,A})\nonumber\\
%&\geq& \mathrm{min_{\ket{j}}}[S(\rho_m)] - S(\rho_{s,A}) =
%\mathcal{D},
%\end{eqnarray}
\begin{eqnarray}
C &=& -\mathrm{Tr}\sum_j\ket{j}\bra{j}\rho_{s,A}\ln
\sum_j\ket{j}\bra{j}\rho_{s,A}\ket{j}\bra{j}- S(\rho_{s,A})\nonumber\\
&=&-\mathrm{Tr}\sum_j\ket{j}\bra{j} \rho_{s,A}\ket{j}\bra{j} \ln
\sum_j\ket{j}\bra{j}\rho_{s,A}\ket{j}\bra{j}\nonumber\\&& -
S(\rho_{s,A})= S(\rho_m) - S(\rho_{s,A})\nonumber\\
&\geq& \mathrm{min_{\ket{j}}}[S(\rho_m)] - S(\rho_{s,A}) =
\mathcal{D},
\end{eqnarray}
where $\mathcal{D}$ is the quantum discord \cite{Modi}, when they assume classicality in terms of only
one sub-system. In their work the discord has the meaning of the distance
between a given state and the closest (system-apparatus) separable state. In other words the lower bound on the confidence is the distance
between the real state and the closest separable state, as one would
expect with projective measurements.

 In the case of a general POVM, one could argue
that the minimization of $C$ over all possible POVMs is equivalent
to a generalization of the definition of Modi's discord  \cite{Modi}. Finally,
we note that there is a correspondence between the estimator state
$\rho_E$ we discussed in terms of the filter equation,  and the
partial trace $\mathrm{Tr}_A(\rho_m)$ over the state constructed
from asymptotic POVM measurements (and the same for the real state
$\rho_R$ and $\mathrm{Tr}_A(\rho_{s,A})$ evolved in the
measurement noise in the filter equation example).

\section{Conclusion}
In many realistic quantum readout architectures the reliability of
the quantum measurement output is an important issue.  In this
article we  discussed how to measure the confidence and the
backaction of a state reconstructed from continuous weak quantum
measurement. As a typical example, we considered a DQD measured by a
nearby QPC. Based on the theory of open quantum systems and the
quantum filter equation method we briefly discussed the trade-off
between measurement confidence and measurement-induced backaction.
We also considered a possible filter equation for ensemble averaged
results. We finished by discussing the case of a general POVM and
how the confidence (when defined as a relative entropy) has a lower
bound related to the quantum discord.

%\input acknowledgement.tex
%\begin{acknowledgements}
%JQY is supported by the National Basic Research Program of China
%Grant No. 2009CB929300, the National Natural Science Foundation of
%China Grant No. 91121015.
\begin{acknowledgements}
We thank Kurt Jacobs for helpful comments. W.C. is supported by the RIKEN FPR Program. Y.O. is partially supported by the Special Postdoctoral Researchers Program, RIKEN. F.N. is partially supported by the ARO, JSPS-RFBR Contract No. 12-02-92100, a Grant-in-Aid for Scientific Research (S), MEXT “Kakenhi on Quantum Cybernetics,” and the JSPS via its FIRST program. J.Q.Y. is partly supported by the National Basic Research Program of China Grant No.2009CB929302, NSFC Grant No.91121015, and MOE Grant No.B06011. X.Y.L. is supported by a JSPS Fellowship.
 \end{acknowledgements}

\appendix
\section{Quantitative characterization of confidence and backaction via
 goal programming }
%%%---Modified-by-YukihiroOta-at-Oct19-2012-start-
%\section{Quantitative characterization of confidence and backaction via goal programming}
\begin{figure}
\setlength{\abovecaptionskip}{7pt}
\includegraphics[width=1\columnwidth]{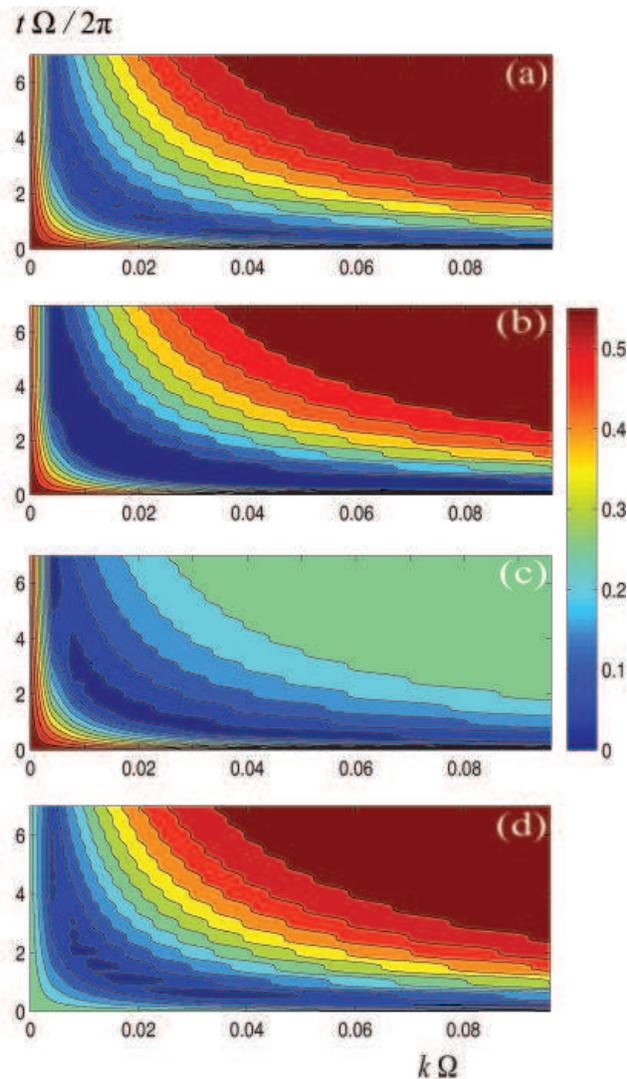}
\caption{(Color online)  Density profile of the optimization function
 $\mathcal{O}$ in the goal programming model for several
 cases: (a) $\eta_1=\eta_2=1$ and $\Delta_C=\Delta_B=0.1$; (b)
 $\eta_1=\eta_2=1$ and $\Delta_C=\Delta_B=0.2$; (c) $\eta_1=1,
 \eta_2=0.5$ and $\Delta_C=\Delta_B=0.1$; (d) $\eta_1=0.5, \eta_2=1$ and
 $\Delta_C=\Delta_B=0.1$.}
\end{figure}

The results in Section IV show a clear trade-off relation
between the confidence and the backaction in the parameter space spanned
by the interaction time $t$ and the measurement strength $k$.
Let us examine this relation via a sophisticated method, goal
programming\,\cite{Goal}.
We formulate our problem setting more specifically; we determine $t$ and
$k$ such that $\mathcal{C} \le \Delta_{C}$ and $\mathcal{B} \le \Delta_{B}$ for given
(small) positive constants $\Delta_{C}$ and $\Delta_{B}$.
The two parameters $\Delta_{C}$ and $\Delta_{B}$ are regarded as,
respectively, admissible confidence error and permissible backaction.
Thus, we can obtain a good measurement scenario to increase the
confidence (i.e., minimize $\mathcal{C}$) while reducing the backaction (i.e.,
minimizing $\mathcal{B}$).
Hereafter, the confidence and the backaction are defined via the relative
entropy, as seen in Eq. (5).

Goal programming \cite{Goal} provides an optimization method to deal
with two (or more than two) conflicting objectives and it is widely used in
mathematics, information theory and engineering.
Instead of finding solutions which absolutely minimize or maximize
objective functions,
the task of goal programming is to find solutions that, if possible,
satisfy a set of goals, or otherwise violate the goals minimally.
This makes the approach more appealing to practical designers, compared
to other optimization methods (e.g., linear programming models).
Let us describe our measurement problem in terms of goal programming:
\begin{eqnarray}
&&
{\rm Minimize}\quad
\mathcal{O}\equiv \eta_{1}\delta_{1}^{+} + \eta_{2}\delta_{2}^{+},
 \nonumber \\
&&
{\rm subject\,\,to}\,
\left\{
\begin{array}{l}
\mathcal{C}(\vec{x})-\delta_{1}^{+}+\delta_{1}^{-} = \Delta_{C} \\
\mathcal{B}(\vec{x})-\delta_{2}^{+}+\delta_{2}^{-} = \Delta_{B} \\
\delta_{1}^{\pm},\delta_{2}^{\pm} \ge 0 \\
\vec{x}=(t\Omega/2\pi, k\Omega) \in \mathcal{M}
\end{array}
\right. .
\label{GP}
\end{eqnarray}
The weight factors $\eta_{1}$ and $\eta_{2}$ are given positive
number, and represent the relative priority of objectives.
If $\eta_{1} > \eta_{2}$, the condition for the
confidence is prior to the one for the backaction, and vice versa.
The condition for the confidence ($\mathcal{C}\le \Delta_{C}$) is reformulated by
the relation
\(
\mathcal{C} +\delta_{1}^{+}-\delta_{1}^{-} = \Delta_{C}
\), with the deviations between the admissible error and the actual
value, $\delta_{1}^{+}$ and $\delta_{1}^{-}$.
When $\mathcal{C} > \Delta_{C}$ ($\mathcal{C} \le \Delta_{C}$),
we set $\delta_{1}^{+}=\mathcal{C}-\Delta_{C}$ and $\delta_{1}^{-}=0$
($\delta_{1}^{+}=0$ and $\delta_{1}^{-}=\Delta_{C}-\mathcal{C}$).
Similarly, we can set $\delta_{2}^{\pm}$ via
$\mathcal{B}-\delta_{2}^{+}+\delta_{2}^{-}=\Delta_{B}$.
The set $\mathcal{M}=\{t\Omega/2\pi,k\Omega\}$ is the family of the
measurement scenarios.
The smaller $\mathcal{O}$, the better performance of the measurement scenario.
The minimum value of $\mathcal{O}$ ($\mathcal{O}=0$) corresponds
to the best solution.

Figure 6 shows the contour profile of $\mathcal{O}$ for various cases,
based on the ensemble-averaged confidence $\mathcal{C}$ and backaction $\mathcal{B}$.
The optimization function $\mathcal{O}$ is a function of the measurement
scenarios: the interaction time $t\Omega/2\pi$ and measurement strength
$k\Omega$.
The other parameters are the same as in Figs.\,4(a) and (b).
In both Figs.\,6(a) and (b), the confidence and the backaction
have equal importance ($\eta_1=\eta_2=1$).
In Fig.6(a) we examine the case that $\Delta_{C}=0.1$ and
$\Delta_{B}=0.1$.
We find that the best solution ($\mathcal{O}=0$) appears in the dark
blue area.
If we relax the restriction to $\Delta_C=\Delta_B=0.2$, we find that
more solutions for the optimization $\mathcal{O}=0$, as seen in Fig.\,6(b).
We also examine the cases when the confidence has a different importance, or weight,
than the backaction, as seen in Figs.\,6(c) and (d).
The solution for the case where the measurement confidence is more important than the backaction ($\eta_1=1, \eta_2=0.5$) is given in Fig. 6(c). The solution for the opposite situation ($\eta_1=0.5, \eta_2=1$) is given in Fig. 6(d).
In the above cases, we have considered a double-criterion goal
programming problem and we find that it is convenient for discussing
the trade-off relation between measurement confidence and measurement
induced backaction.

%\begin{figure}
%\setlength{\abovecaptionskip}{7pt}
%\centerline{\scalebox{0.7}[0.61]{\includegraphics{fig4.eps}}}
% \caption{(Color online) Ensemble-averaged (a) confidence $C$ and
% (b) backaction $B$ in terms of the relative entropy between system and
% estimator states, as a function of the interaction time $t$ and measurement
% strength $\kappa$.  Both figures
% are derived using the ensemble-averaged equation of motion (\ref{ME2}).
% (c) and (d) show a comparison between the probability of occupation of
% the left dot $P_L(t)$ derived from averaging Eq. [\ref{ME}] over many
% realizations (dashed blue lines) to that
% given by solving Eq. [\ref{ME2}].}
%\end{figure}

\end{document}